\documentclass[prl,twocolumn,showpacs]{revtex4-1}
\usepackage{amsmath,bm,graphicx,hyperref}

\newcommand\grad{\bm{\nabla}}
\newcommand\+{\dagger}

\renewcommand\>{\rangle}
\newcommand\z{{\bm{z}}}
\renewcommand\k{{\bm{k}}}
\newcommand\p{{\bm{p}}}
\newcommand\q{{\bm{q}}}
\renewcommand\r{{\bm{r}}}
\newcommand\ek{\epsilon_\k}
\newcommand\ep{\epsilon_\p}
\newcommand\eq{\epsilon_\q}
\newcommand\kB{k_\mathrm{B}}
\newcommand\kF{k_\mathrm{F}}
\newcommand\EF{E_\mathrm{F}}
\newcommand\Es{E_\mathrm{s}}
\newcommand\Et{E_\mathrm{t}}
\newcommand\TF{T_\mathrm{F}}
\newcommand\TK{T_\mathrm{K}}
\newcommand\FS{\mathrm{FS}}
\renewcommand\Re{\mathrm{Re}}
\renewcommand\Im{\mathrm{Im}}
\newcommand\D{\mathcal{D}}
\newcommand\E{\mathcal{E}}
\newcommand\T{\mathcal{T}}
\newcommand\Z{\mathcal{Z}}
\newcommand\sect[1]{\subsection{#1}}

\begin{document}

\title{SU(3) orbital Kondo effect with ultracold atoms}

\author{Yusuke Nishida}
\affiliation{Department of Physics, Tokyo Institute of Technology,
Ookayama, Meguro, Tokyo 152-8551, Japan}

\date{August 2013}

\begin{abstract}
 We propose a simple but novel scheme to realize the Kondo effect with
 ultracold atoms.  Our system consists of a Fermi sea of spinless
 fermions interacting with an impurity atom of different species which
 is confined by an isotropic potential.  The interspecies attraction can
 be tuned with an $s$-wave Feshbach resonance so that the impurity atom
 and a spinless fermion form a bound dimer that occupies a
 threefold-degenerate $p$ orbital of the confinement potential.
 Many-body scatterings of this dimer and surrounding spinless fermions
 occur with exchanging their angular momenta and thus exhibit the SU(3)
 orbital Kondo effect.  The associated Kondo temperature has a universal
 leading exponent given by $\TK\propto\exp[-\pi/(3a_p\kF^3)]$ that
 depends only on an effective $p$-wave scattering volume $a_p$ and a
 Fermi wave vector $\kF$.  We also elucidate a Kondo singlet formation
 at zero temperature and an anisotropic interdimer interaction mediated
 by surrounding spinless fermions.  The Kondo effect thus realized in
 ultracold atom experiments may be observed as an increasing atom loss
 by lowering the temperature or with radio-frequency spectroscopy.  Our
 scheme and its extension to a dense Kondo lattice will be useful to
 develop new insights into yet unresolved aspects of Kondo physics.
\end{abstract}


\maketitle

\sect{Introduction}
The Kondo effect plays fundamental roles in a wide range of condensed
matter physics.  The Kondo effect arises from coherent many-body
scatterings between a localized magnetic impurity and a surrounding
Fermi sea of itinerant electrons with the exchange of their
spins~\cite{Hewson:1993,Coleman:2007}.  While the Kondo effect was
originally discovered in explaining an anomalous electrical resistivity
minimum in dilute magnetic alloys~\cite{Kondo:1964}, it has been
observed also in artificial nanosystems such as quantum
dots~\cite{Goldhaber-Gordon:1998,Cronenwett:1998,Schmid:1998}, carbon
nanotubes~\cite{Nygard:2000,Buitelaar:2002}, and individual
molecules~\cite{Park:2002,Liang:2002}.  Furthermore, the role of the
spin can be replaced by other degrees of freedom such as an orbital
quantum number~\cite{Cox:1998}, which leads to the orbital Kondo
effect~\cite{Kolesnychenko:2002,Jarillo-Herrero:2005}.  In spite of the
fact that the Kondo effect is well known and widely studied, there are
yet unresolved issues, for example, regarding the formation of the Kondo
screening cloud and its dynamics~\cite{Kouwenhoven:2001}.

Ultracold atoms may be useful to develop new insights into this old but
still actively studied phenomenon.  This is because a highly tunable
clean system that is suitable to study a desired phenomenon can be
created at will by using optical lattices and Feshbach resonances.  This
scheme called ``quantum simulation'' has been successfully implemented
to study a rich variety of strongly correlated many-body
phenomena~\cite{Bloch:2012}.  Regarding the realization of the Kondo
effect, essential ingredients are (i) degenerate degrees of freedom that
can be exchanged between a localized impurity and surrounding fermions
and (ii) a mechanism that suppresses a double occupancy of the localized
state.  While both ingredients are naturally incorporated in condensed
matter systems by the spin and Coulomb repulsion of charged electrons,
they are in general difficult to achieve simultaneously with neutral
atoms.  Accordingly, previously proposed theoretical schemes to realize
the Kondo effect with ultracold
atoms~\cite{Recati:2002,Falco:2004,Paredes:2005,Duan:2004,Gorshkov:2010,Lal:2010}
are somewhat involved and demanding and thus the experimental
realization is still lacking.

In this Letter, we propose a simple but novel scheme to realize the
Kondo effect with ultracold atoms.  Our system consists of a Fermi sea
of spinless fermions of species $A$ interacting with an impurity atom of
different species $B$ which is loaded into a ground state of an
isotropic potential $V_B(r)$:
\begin{align}\label{eq:hamiltonian}
 H &= \int d\r\,\Psi_A^\+(\r)
 \!\left[-\frac{\hbar^2\grad^2}{2m_A}-\EF\right]\!\Psi_A(\r) \notag\\
 &\ \ + \int d\r\,\Psi_B^\+(\r)
 \!\left[-\frac{\hbar^2\grad^2}{2m_B}+V_B(r)\right]\!\Psi_B(\r) \notag\\
 &\ \ + U_{AB}\int d\r\,\Psi_A^\+(\r)\Psi_B^\+(\r)\Psi_B(\r)\Psi_A(\r).
\end{align}
By tuning the interspecies attraction $U_{AB}<0$ with an $s$-wave
Feshbach resonance, the impurity atom and a spinless fermion can form a
bound dimer that occupies a threefold-degenerate $p$ orbital of the
confinement potential~\cite{Nishida:2010}.  It is important to note that
the impurity atom cannot bind two spinless fermions simultaneously
because of the Pauli exclusion principle.  Many-body scatterings of this
dimer and surrounding spinless fermions occur with exchanging their
angular momenta and thus exhibit the SU(3) orbital Kondo effect.  In
what follows, we give a detailed account of how the Kondo effect emerges
in our simple system and elucidate its fundamental properties such as a
universal leading exponent of the Kondo temperature, a Kondo singlet
formation at zero temperature, an anisotropic interdimer interaction
mediated by surrounding spinless fermions, and possible experimental
signatures.  Below we denote a kinetic energy of spinless fermions by
$\ek=\hbar^2\k^2/(2m_A)$ and set $\hbar=1$.

\sect{Few-body scattering and effective theories}
Before working on the many-body Kondo effect, we first need to elucidate
the few-body scattering properties of our system (\ref{eq:hamiltonian}).
A two-body scattering of an $A$ atom with a confined $B$ atom was
studied in Ref.~\cite{Nishida:2010}.  There it was found that by tuning
the interspecies attraction with an $s$-wave Feshbach resonance, a
$p$-wave scattering resonance can be induced.  An intuitive picture for
this seemingly counterintuitive result is actually
simple~\cite{Nishida:2010,Lamporesi:2010,Massignan:2006}:  A Feshbach
molecule formed by the $A$ and $B$ atoms is subject to the confinement
potential and thus its center-of-mass energy is quantized.  By
decreasing its internal energy with the interspecies attraction, the
total energy of the $AB$ molecule in the $p$ orbital intersects the
scattering threshold of the $A$ and $B$ atoms.  At this point, the
$p$-wave scattering resonance is induced and low-energy physics in its
vicinity is described by a two-channel Hamiltonian:
\begin{align}\label{eq:2-channel}
 & H_\mathrm{2ch}
 = \int\!\frac{d\k}{(2\pi)^3}\,\ek\,\psi_A^\+(\k)\psi_A(\k)
 + \epsilon_{AB}\sum_\mu\phi_\mu^\+\phi_\mu \notag\\
 &\quad + g\sum_\mu\int^\Lambda\!\!\frac{d\k}{(2\pi)^3}\,k_\mu
 \!\left[\psi_A^\+(\k)\psi_B^\+\phi_\mu + \phi_\mu^\+\psi_B\psi_A(\k)\right]\!.
\end{align}
Here $\psi_B^\+$ is a creation operator of the impurity $B$ atom in the
ground state of the isotropic potential whose energy is chosen to be
zero, while $\phi_\mu^\+$ is that of the $AB$ molecule in one of the
threefold-degenerate $p$ orbitals labeled by $\mu=x,y,z$.  Because of
the angular momentum conservation, this $AB$ molecule can couple with
the impurity $B$ atom and an $A$ atom only with the same orbital quantum
number $\mu$, which is created by
$\psi_{A\mu}^\+(k)\equiv\int\!d\Omega\,\hat{k}_\mu\psi_A^\+(\k)$.  The
particle number operators of the localized $B$ atom and $AB$ molecule
are constrained by
\begin{align}\label{eq:constraint}
 N_B = \psi_B^\+\psi_B+\sum_\mu\phi_\mu^\+\phi_\mu = 1
\end{align}
because only one $B$ atom is confined.  Therefore, the above two-channel
Hamiltonian corresponds to the infinite-$U$ Anderson impurity model in
the slave-particle representation~\cite{Barnes:1976,Coleman:1984}.  We
also note that the same Hamiltonian (\ref{eq:2-channel}) can be realized
with an interspecies $p$-wave Feshbach resonance, which is referred to
as a $p$-wave Fano-Anderson model in Ref.~\cite{Gurarie:2007}.

The low-energy effective description (\ref{eq:2-channel}) is valid well
below a ultraviolet cutoff $\Lambda\sim\sqrt{m_B\omega_\mathrm{ho}}$,
which is a momentum scale set by an inverse characteristic extent of the
confined $B$ atom.  The remaining two couplings $\epsilon_{AB}$ and $g$
can be related to physical parameters to characterize a low-energy
$p$-wave scattering, namely, a scattering volume $a_p$ and an effective
momentum $r_p$.  By matching the two-body scattering amplitude computed
from Eq.~(\ref{eq:2-channel}) with the standard formula, we find
$-\frac{\epsilon_{AB}}{g^2}+\frac{m_A\Lambda^3}{9\pi^2}=\frac{m_A}{6\pi}\frac1{a_p}$
and
$\frac1{2m_Ag^2}+\frac{m_A\Lambda}{3\pi^2}=-\frac{m_A}{6\pi}\frac{r_p}2>0$,
respectively.  Information on $V_B(r)$ and $U_{AB}$ in the original
Hamiltonian (\ref{eq:hamiltonian}) is now encoded into $a_p$ and $r_p$,
which have been determined as functions of an $s$-wave scattering length
in the case of a harmonic potential
$V_B(r)=m_B\omega_\mathrm{ho}^2r^2/2$~\cite{Nishida:2010}.  (It is
worthwhile to note that, unlike an $s$-wave scattering resonance, the
cutoff $\Lambda$ cannot be sent to infinite with $a_p$ and $r_p$ fixed
because $|r_p|>4\Lambda/\pi$.)  By using $a_p$ and $r_p$, the
renormalized Green's function of the $AB$ molecule at energy $E$ is
expressed as $g^2\<T\phi_\nu\phi_\mu^\+\>=i\delta_{\mu\nu}\D(E)$, where
\begin{align}\label{eq:propagator}
 \D(E) = \frac{6\pi}{m_A}\frac1{\frac1{a_p}-\frac{|r_p|}2\kappa^2
 +\frac2\pi\kappa^3\arctan\!\left(\frac\Lambda\kappa\right)}
\end{align}
with $\kappa\equiv\sqrt{-2m_AE-i0^+}$.  When the effective $p$-wave
scattering volume $a_p$ is positive, Eq.~(\ref{eq:propagator}) develops
a pole at $E=\E<0$ with a corresponding residue $\Z$.  This means the
existence of threefold-degenerate bound dimer states that carry an
orbital angular momentum $\ell=1$.  Because of the constraint
(\ref{eq:constraint}), only one of them can be occupied at one time.

\begin{figure}[t]
 \includegraphics[width=\columnwidth,clip]{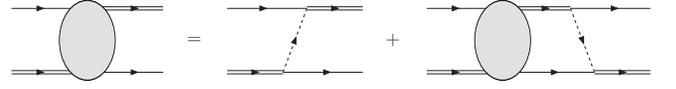}
 \caption{Atom-dimer scattering $\T$ matrix in the vacuum represented by
 the blob.  Solid, dotted, and double lines correspond to Green's
 functions of the $A$ atom, $B$ atom, and $AB$ molecule, respectively.
 \label{fig:t_vacuum}}
\end{figure}

We then study a scattering of this bound dimer with another $A$ atom
which is now a three-body problem.  The corresponding scattering
$\T$ matrix at energy $E$ solves an integral equation depicted in
Fig.~\ref{fig:t_vacuum}:
\begin{align}\label{eq:integral_eq}
 & \T_{\mu\nu}(P;Q) = -\frac{q_\mu p_\nu\Z}{E-p_0-q_0+i0^+} \notag\\
 &\quad - \sum_\lambda\int^\Lambda\!\!\frac{d\k}{(2\pi)^3}\,\T_{\mu\lambda}(P;K)\,
 \frac{q_\lambda k_\nu\D(E-k_0)}{E-k_0-q_0+i0^+}\,\bigg|_{k_0=\ek}.
\end{align}
Here $P=(p_0,\p)$ [$Q=(q_0,\q)$] is a set of energy and momentum of the
incoming (outgoing) $A$ atom and $\mu$ ($\nu$) labels a $p$ orbital that
is initially (finally) occupied by the bound dimer.  The atom-dimer
scattering $\T$ matrix at their scattering threshold is obtained from
$\T_{\mu\nu}(P;Q)$ by setting $p_0=q_0=0$ and $E=\E$.  The resulting
quantity $t_{\mu\nu}(\p;\q)\equiv\T_{\mu\nu}(0,\p;0,\q)|_{E=\E}$ is in
general cutoff dependent and thus not universal, but we find that it
becomes universal in the vicinity of the $p$-wave scattering resonance:
\begin{align}\label{eq:t_vacuum}
 \lim_{a_p\to+\infty}t_{\mu\nu}(\p;\q) = \frac{6\pi a_p}{m_A}q_\mu p_\nu
\end{align}
with $g<\infty$ and $\Lambda$ fixed.  Technically, this is because the
source term in Eq.~(\ref{eq:integral_eq}) dominates over the integral
term, which means that the Born approximation turns out to be exact for
this particular quantity sufficiently close to the resonance.  A similar
universality was found in the corresponding atom-dimer scattering
problem without the confinement potential~\cite{Jona-Lasinio:2008}.  We
also confirmed by numerically solving the integral equation
(\ref{eq:integral_eq}) that three-body bound states do not exist in the
same limit as in Eq.~(\ref{eq:t_vacuum}); i.e., the impurity atom cannot
bind two spinless fermions simultaneously.

We finally arrive at a stage to write down an effective Hamiltonian
that describes the above low-energy scattering of an $A$ atom with a
bound dimer in the vicinity of the $p$-wave scattering resonance
$a_p\Lambda^3\gg1$:
\begin{align}\label{eq:atom-dimer}
 & H_\mathrm{ad}
 = \int\!\frac{d\k}{(2\pi)^3}\,\ek\,\psi_A^\+(\k)\psi_A(\k) \notag\\
 &\quad + \sum_{\mu,\nu}\int^{\Lambda'}\!\frac{d\p\,d\q}{(2\pi)^6}\,
 \psi_A^\+(\q)\psi_A(\p)\bigl[v\,q_\mu p_\nu
 + v'\delta_{\mu\nu}\q\cdot\p\bigr]\phi_\nu^\+\phi_\mu.
\end{align}
Here the two couplings
$v=\frac{6\pi a_p}{m_A}/\bigl[1-\bigl(\frac{2a_p\Lambda'^3}{3\pi}\bigr)^2\bigr]$
and $v'=\frac{2a_p\Lambda'^3}{3\pi}v$ are determined so that the
zero-energy atom-dimer scattering $\T$ matrix $t_{\mu\nu}(\p;\q)$
obtained in Eq.~(\ref{eq:t_vacuum}) is correctly reproduced from
Eq.~(\ref{eq:atom-dimer}).  This low-energy effective description is
valid well below the new ultraviolet cutoff
$\Lambda'\sim1/\sqrt{a_p|r_p|}$, which is now a momentum scale set by an
inverse characteristic extent of the bound dimer.  The atom-dimer
Hamiltonian (\ref{eq:atom-dimer}) reduced from Eq.~(\ref{eq:2-channel})
corresponds to the Kondo or Coqblin-Schrieffer model reduced from the
Anderson impurity model~\cite{Schrieffer:1966,Coqblin:1969} and thus
serves as a basis for the many-body Kondo effect.

\begin{figure}[t]
 \includegraphics[width=0.95\columnwidth,clip]{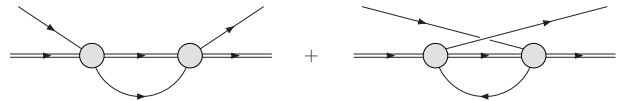}
 \caption{Subleading contributions to the atom-dimer scattering
 $\T$ matrix in the presence of a Fermi sea of spinless fermions.  See
 Fig.~\ref{fig:t_vacuum} for the meanings of the symbols.
 \label{fig:t_medium}}
\end{figure}

\sect{Many-body physics and the Kondo effect}
We are now prepared to work on the Kondo effect.  We apply the
atom-dimer Hamiltonian (\ref{eq:atom-dimer}) to a many-body system of
spinless fermions with a Fermi energy $\EF\equiv\kF^2/2m_A\equiv\kB\TF$
at a temperature $T$ and study their scatterings with the localized
dimer.  We consider that the incoming (outgoing) fermion has a momentum
$\p=\kF\hat\p$ ($\q=\kF\hat\q$) and the localized dimer initially
(finally) occupies a $p$ orbital labeled by $\mu$ ($\nu$) and compute
the corresponding scattering $\T$ matrix $t_{\mu\nu}(\p;\q)$ in a
perturbative expansion over $a_p\kF^3,\,a_p\kF^2\Lambda'\ll1$ assuming a
dilute system.  The leading $O(a_p)$ term is simply the one obtained in
Eq.~(\ref{eq:t_vacuum}) and thus does not involve the many-body effect.
There are two distinct contributions at subleading $O(a_p^2)$ depending
on whether the intermediate scattering state is a particle or a hole,
which is depicted by the left or right diagram in
Fig.~\ref{fig:t_medium}, respectively.  By summing all contributions up
to $O(a_p^2)$, the zero-energy atom-dimer scattering $\T$ matrix in the
low-temperature limit $T\ll\TF$ is found to be
\begin{align}\label{eq:t_medium}
 & \Re\,t_{\mu\nu}(\p;\q)
 = \left(q_\mu p_\nu - \frac{\delta_{\mu\nu}}3\q\cdot\p\right)v(T) \notag\\
 &\quad + \frac{\delta_{\mu\nu}}3\q\cdot\p
 \left(v_0 - v_0^2\frac{m_A\kF^2}{\pi^2}\Lambda'\right) + O(a_p^3),
\end{align}
where $v_0=6\pi a_p/m_A>0$ is the atom-dimer coupling in the vacuum [see
Eq.~(\ref{eq:t_vacuum})].

The first (second) term in Eq.~(\ref{eq:t_medium}) corresponds to a
scattering process in which orbital quantum numbers are (not) exchanged
between the spinless fermion and the localized dimer.  While the second
term is temperature independent, the first term depends on the
temperature through
\begin{align}\label{eq:coupling}
 v(T) = v_0 + v_0^2\frac{m_A\kF^3}{2\pi^2}
 \left[\,\ln\!\left(\frac{\TF}T\right)+\mathrm{const}\right] + O(a_p^3).
\end{align}
This effective atom-dimer coupling exhibits the logarithmic growth by
lowering the temperature which is the hallmark of the Kondo
effect~\cite{Kondo:1964}.  In particular, this is an SU(3) orbital Kondo
effect because our low-energy effective Hamiltonian (\ref{eq:2-channel})
or (\ref{eq:atom-dimer}) is invariant under an SU(3) rotation of
$p$-orbital indices of $\phi_\mu$ and $\psi_{A\mu}(k)$.  Furthermore,
the leading logarithmic singularities can be summed up to all orders
with the aid of the renormalization group equation:
$\frac{dv(T)}{d\ln(\TF/T)}=\frac{m_A\kF^3}{2\pi^2}v^2(T)$, which is
solved by
$v(T)\simeq v_0/\bigl[1-v_0\frac{m_A\kF^3}{2\pi^2}\ln\bigl(\frac{\TF}T\bigr)\bigr]$.
This renormalized effective coupling diverges at a Kondo temperature
given by
\begin{align}\label{eq:temperature}
 \TK\propto\TF\exp\!\left(-\frac\pi{3a_p\kF^3}\right).
\end{align}
This expression is valid in the vicinity of the $p$-wave scattering
resonance $a_p\Lambda^3\gg1$ for a sufficiently dilute system
$a_p\kF^3,\,a_p|r_p|\kF^2\ll1$.  Remarkably, we find that the Kondo
temperature has a universal leading exponent that depends on details of
the original Hamiltonian (\ref{eq:hamiltonian}) only through the
effective $p$-wave scattering volume $a_p$ and the Fermi wave vector
$\kF$.

The rate at which spinless fermions on the Fermi surface are scattered
by localized dimers with a density $n_\mathrm{d}$ is obtained from the
standard formula as
\begin{align}\label{eq:rate}
 \frac1{\tau_\mathrm{ad}} &= \frac{n_\mathrm{d}}3\sum_{\mu,\nu}
 \int\!\frac{d\q}{(2\pi)^3}\,2\pi\,\delta(\ep-\eq)\,|t_{\mu\nu}(\p;\q)|^2 \notag\\
 &= \frac{8m_An_\mathrm{d}\kF^5}{27\pi}v^2(T) + \mathrm{const},
\end{align}
which grows logarithmically according to Eq.~(\ref{eq:coupling}).  An
atom-dimer scattering in ultracold atom experiments typically leads to a
loss of atoms because the dimer can decay into a deeper bound
state~\cite{Kohler:2006,Chin:2010} or a lower $s$ orbital of the
confinement potential in our scheme.  Therefore, the logarithmic growth
of the atom-dimer scattering rate (\ref{eq:rate}) caused by the Kondo
effect may be observed as an increasing atom loss by lowering the
temperature, which is a direct analog of the celebrated increasing
electrical resistivity in dilute magnetic alloys~\cite{de-Haas:1934}.

The divergence of the effective atom-dimer coupling (\ref{eq:coupling})
means the breakdown of the perturbation theory below the Kondo
temperature $T\lesssim\TK$.  Here the localized dimer interacts so
strongly with surrounding spinless fermions that its angular momentum is
effectively screened by the so-called Kondo screening cloud.
Accordingly, the ground state becomes a singlet state instead of a
naively expected triplet state.  This nonperturbative physics can be
captured by the following variational wave functions corresponding to
the nondegenerate singlet state:
\begin{align}\label{eq:singlet}
 & |\mathrm{s}\> = \left[\alpha\,\psi_B^\+
 + \sum_\mu\int^{\kF}\!\!\!\frac{d\p}{(2\pi)^3}\,
 \beta_\mu(\p)\phi_\mu^\+\psi_A(\p)\right. \notag\\
 &\quad + \left.\int^{\kF}\!\!\!\frac{d\p}{(2\pi)^3}\int_{\kF}\!\frac{d\q}{(2\pi)^3}\,
 \gamma(\p,\q)\psi_B^\+\psi_A^\+(\q)\psi_A(\p)\right]|\FS\>
\end{align}
and the threefold-degenerate triplet state ($\mu=x,y,z$):
\begin{align}\label{eq:triplet}
 |\mathrm{t}\>_\mu = \left[\bar\alpha\,\phi_\mu^\+
 + \int_{\kF}\!\frac{d\q}{(2\pi)^3}\,
 \bar\beta_\mu(\q)\psi_B^\+\psi_A^\+(\q)\right]|\FS\>,
\end{align}
where $|\FS\>$ represents the Fermi sea of $A$ atoms.  The singlet state
consists of a localized $B$ atom dressed by a particle-hole excitation
of $A$ atoms which hybridizes with an $AB$ molecule by absorbing an $A$
atom from the Fermi sea.  On the other hand, the triplet state consists
of a localized $AB$ molecule which hybridizes with a $B$ atom by
emitting an $A$ atom to the Fermi sea.  These two states have the same
particle number $N_B=1$ in compliance with the constraint
(\ref{eq:constraint}) but have different orbital angular momenta,
$\ell=0$ and $1$, respectively.

By minimizing the expectation values of the two-channel Hamiltonian
(\ref{eq:2-channel}) with respect to the variational parameters in
Eqs.~(\ref{eq:singlet}) and (\ref{eq:triplet}), we find that the
energies of the singlet state $\Es$ and the triplet state $\Et$ apart
from the energy of the Fermi sea solve $G^{-1}(\Es)=0$ and
$D^{-1}(\Et+\EF)=0$, respectively, where
\begin{align}
 D^{-1}(E) = \D^{-1}(E)
 + \frac13\int\!\frac{d\q}{(2\pi)^3}\frac{q^2\theta(\kF-q)}{E-\eq}
\end{align}
is the inverse Green's function of the $AB$ molecule and
\begin{align}
 G^{-1}(E) = E - \int\!\frac{d\p}{(2\pi)^3}\,p^2\theta(\kF-p)D(E+\ep)
\end{align}
is that of the $B$ atom in the presence of the Fermi sea of $A$ atoms.
In the resonance and dilute limits
$a_p\kF^3,\,a_p|r_p|\kF^2\ll1\ll a_p\Lambda^3$ considered above [see
below Eq.~(\ref{eq:temperature})], the energy of the triplet state
approaches that of the bound dimer $\Et=-\EF-\frac1{m_Aa_p|r_p|}$.
Although this triplet state is already deeply bound with respect to the
Fermi energy, the singlet state has an even lower energy given by
\begin{align}\label{eq:binding}
 \Es &= \Et - \EF\exp\!\left(-\frac\pi{3a_p\kF^3}
 -\frac{\pi|r_p|}{6\kF}-\frac{8-6\ln2}3\right) \notag\\
 &\simeq \Et - \kB\TK,
\end{align}
where the same universal leading exponent as the Kondo temperature
(\ref{eq:temperature}) appears in the binding energy.  We also confirmed
numerically within the variational wave functions (\ref{eq:singlet}) and
(\ref{eq:triplet}) that the ground state is always singlet and thus the
transition to the triplet state does not take place as a consequence of
the Kondo effect.  Such a smooth crossover at zero temperature is in
contrast to the polaron-molecule transition in the corresponding
impurity problem without the confinement potential which was found both
for $s$-wave and $p$-wave Feshbach
resonances~\cite{Chevy:2010,Levinsen:2012}.  We finally note that the
spectral density function of the impurity $B$ atom is obtained from
$\rho_B(\omega)=-\frac1\pi\,\Im\,G(\omega+i0^+)$, which exhibits a
$\delta$-function peak at $\omega=\Es$ and a continuum above
$\omega=\Et$.  This quantity can be measured in ultracold atom
experiments with the radio-frequency
spectroscopy~\cite{Schirotzek:2009,Kohstall:2012,Knap:2012}, and,
accordingly, the binding energy of the Kondo singlet (\ref{eq:binding})
as well.

\sect{Concluding remarks}
In this Letter, we proposed and elaborated a simple but novel scheme to
realize the SU(3) orbital Kondo effect with ultracold atoms.  In
addition to its simplicity, the advantage of our system
(\ref{eq:hamiltonian}) also lies in its versatility.  For example, by
deforming the isotropic confinement potential $V_B(r)$, the $p$-orbital
degeneracy can be reduced from threefold to twofold, which now leads to
an SU(2) orbital Kondo effect.  Moreover, by further increasing the
interspecies attraction $U_{AB}$, an $\ell$\,th partial-wave scattering
resonance can be induced~\cite{Nishida:2010}, where an SU($2\ell+1$)
orbital Kondo effect will emerge in the same system
(\ref{eq:hamiltonian}).  The Kondo effect thus realized in our scheme
causes a logarithmic growth of the atom-dimer scattering rate, which may
be observed in ultracold atom experiments as an increasing atom loss by
lowering the temperature.  This hallmark is a direct analog of the
celebrated increasing electrical resistivity in dilute magnetic
alloys~\cite{Kondo:1964,de-Haas:1934}.  Moreover, the spectral density
function of the Kondo singlet and its binding energy can be measured in
our scheme with the radio-frequency spectroscopy, which will be useful
to develop further insights into the Kondo physics.

\begin{figure}[t]
 \includegraphics[width=0.35\columnwidth,clip]{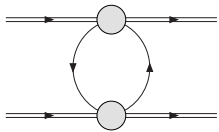}
 \caption{Leading interdimer interaction mediated by a surrounding Fermi
 sea of spinless fermions.  See Fig.~\ref{fig:t_vacuum} for the meanings
 of the symbols.  \label{fig:rkky}}
\end{figure}

Our scheme can be easily extended to a dense Kondo lattice where
localized impurities are placed periodically and a competition between
the Kondo effect and the Ruderman-Kittel-Kasuya-Yosida (RKKY)
interaction is expected to lead to a rich phase diagram with a quantum
critical point~\cite{Coleman:2007,Si:2010}.  In particular, the RKKY
interaction in our system is unusual because orbital indices of
localized impurities are correlated with a direction in which they are
separated~\cite{Coqblin:1969}.  This feature can be seen by considering
two bound dimers localized at $\r$ and $\r'$ with a separation
$\r-\r'=R\hat\z$ chosen in the $z$ direction.  Here surrounding spinless
fermions mediate an effective interaction between the two localized
dimers, which can be computed from the atom-dimer Hamiltonian
(\ref{eq:atom-dimer}) in a perturbative expansion over $a_p\kF^3\ll1$.
The leading $O(a_p^2)$ term is depicted in Fig.~\ref{fig:rkky}, which at
a large separation $\kF R\gg1$ is found to be universal, i.e.,
independent of the cutoff $\Lambda'$, and given by
\begin{align}
 H_\mathrm{RKKY} &= \frac{9a_p^2\kF^5\cos(2\kF R)}{2\pi m_AR^3}
 \phi_z^\+(\r)\phi_z(\r)\phi_z^\+(\r')\phi_z(\r') \notag\\
 &\ \ + O(R^{-4}) + O(a_p^3),
\end{align}
where $p$-orbital indices of the localized dimers are locked to the
direction of their separation.  Because of the geometrical frustration,
such a strongly anisotropic interdimer interaction may lead to an
interesting many-body physics of the dense Kondo lattice realized in our
scheme, which is to be studied in the future.


\begin{thebibliography}{99}

\bibitem{Hewson:1993}
  A.~C.~Hewson,
  {\it The Kondo Problem to Heavy Fermions}
  (Cambridge University Press, Cambridge, England, 1993).

\bibitem{Coleman:2007}
  P.~Coleman,
  in {\it Handbook of Magnetism and Advanced Magnetic Materials,}
  edited by H.~Kronmueller and S.~Parkin
  (John Wiley \& Sons, New York, 2007).

\bibitem{Kondo:1964}
  J.~Kondo,
  Prog.\ Theor.\ Phys.\ {\bf 32}, 37 (1964).

\bibitem{Goldhaber-Gordon:1998}
  D.~Goldhaber-Gordon, H.~Shtrikman, D.~Mahalu, D.~Abusch-Magder, U.~Meirav, and M.~A.~Kastner,
  Nature (London) {\bf 391}, 156 (1998).

\bibitem{Cronenwett:1998}
  S.~M.~Cronenwett, T.~H.~Oosterkamp, and L.~P.~Kouwenhoven,
  Science {\bf 281}, 540 (1998).

\bibitem{Schmid:1998}
  J.~Schmid, J.~Weis, K.~Eberl, and K.~von~Klitzing,
  Physica (Amsterdam) B {\bf 256--258}, 182 (1998).

\bibitem{Nygard:2000}
  J.~Nyg$\mathring{\mathrm{a}}$rd, D.~H.~Cobden, and P.~E.~Lindelof,
  Nature (London) {\bf 408}, 342 (2000).

\bibitem{Buitelaar:2002}
  M.~R.~Buitelaar, A.~Bachtold, T.~Nussbaumer, M.~Iqbal, and C.~Sch\"onenberger,
  Phys.\ Rev.\ Lett.\ {\bf 88}, 156801 (2002).

\bibitem{Park:2002}
  J.~Park, A.~N.~Pasupathy, J.~I.~Goldsmith, C.~Chang, Y.~Yaish, J.~R.~Petta, M.~Rinkoski, J.~P.~Sethna, H.~D.~Abru\~na, P.~L.~McEuen, and D.~C.~Ralph,
  Nature (London) {\bf 417}, 722 (2002).

\bibitem{Liang:2002}
  W.~Liang, M.~P.~Shores, M.~Bockrath, J.~R.~Long, and H.~Park,
  Nature (London) {\bf 417}, 725 (2002).

\bibitem{Cox:1998}
  D.~L.~Cox and A.~Zawadowski,
  Adv.\ Phys.\ {\bf 47}, 599 (1998).

\bibitem{Kolesnychenko:2002}
  O.~Yu.~Kolesnychenko, R.~de~Kort, M.~I.~Katsnelson, A.~I.~Lichtenstein, and H.~van~Kempen,
  Nature (London) {\bf 415}, 507 (2002).

\bibitem{Jarillo-Herrero:2005}
  P.~Jarillo-Herrero, J.~Kong, H.~S.~J.~van~der~Zant, C.~Dekker, L.~P.~Kouwenhoven, and S.~De~Franceschi,
  Nature (London) {\bf 434}, 484 (2005).

\bibitem{Kouwenhoven:2001}
  L.~Kouwenhoven and L.~Glazman,
  Phys.\ World {\bf 14}, 33 (2001).

\bibitem{Bloch:2012}
  I.~Bloch, J.~Dalibard, and S.~Nascimb\`ene,
  Nat.\ Phys.\ {\bf 8}, 267 (2012).

\bibitem{Recati:2002}
  A.~Recati, P.~O.~Fedichev, W.~Zwerger, J.~von~Delft, and P.~Zoller,
  arXiv:cond-mat/0212413.

\bibitem{Falco:2004}
  G.~M.~Falco, R.~A.~Duine, and H.~T.~C.~Stoof,
  Phys.\ Rev.\ Lett.\ {\bf 92}, 140402 (2004).

\bibitem{Paredes:2005}
  B.~Paredes, C.~Tejedor, and J.~I.~Cirac,
  Phys.\ Rev.\ A {\bf 71}, 063608 (2005).

\bibitem{Duan:2004}
  L.-M.~Duan,
  Europhys.\ Lett.\ {\bf 67}, 721 (2004).

\bibitem{Gorshkov:2010}
  A.~V.~Gorshkov, M.~Hermele, V.~Gurarie, C.~Xu, P.~S.~Julienne, J.~Ye, P.~Zoller, E.~Demler, M.~D.~Lukin, and A.~M.~Rey,
  Nat.\ Phys.\ {\bf 6}, 289 (2010).

\bibitem{Lal:2010}
  S.~Lal, S.~Gopalakrishnan, and P.~M.~Goldbart,
  Phys.\ Rev.\ B {\bf 81}, 245314 (2010).

\bibitem{Nishida:2010}
  Y.~Nishida and S.~Tan,
  Phys.\ Rev.\ A {\bf 82}, 062713 (2010).

\bibitem{Lamporesi:2010}
  G.~Lamporesi, J.~Catani, G.~Barontini, Y.~Nishida, M.~Inguscio, and F.~Minardi,
  Phys.\ Rev.\ Lett.\ {\bf 104}, 153202 (2010).

\bibitem{Massignan:2006}
  P.~Massignan and Y.~Castin,
  Phys.\ Rev.\ A {\bf 74}, 013616 (2006).

\bibitem{Barnes:1976}
  S.~E.~Barnes,
  J.\ Phys.\ F {\bf 6}, 1375 (1976);
  {\bf 7}, 2637 (1977).

\bibitem{Coleman:1984}
  P.~Coleman,
  Phys.\ Rev.\ B {\bf 29}, 3035 (1984).

\bibitem{Gurarie:2007}
  V.~Gurarie and L.~Radzihovsky,
  Ann.\ Phys.\ (Amsterdam) {\bf 322}, 2 (2007).

\bibitem{Jona-Lasinio:2008}
  M.~Jona-Lasinio, L.~Pricoupenko, and Y.~Castin,
  Phys.\ Rev.\ A {\bf 77}, 043611 (2008).

\bibitem{Schrieffer:1966}
  J.~R.~Schrieffer and P.~A.~Wolff,
  Phys.\ Rev.\ {\bf 149}, 491 (1966).

\bibitem{Coqblin:1969}
  B.~Coqblin and J.~R.~Schrieffer,
  Phys.\ Rev.\ {\bf 185}, 847 (1969).

\bibitem{Kohler:2006}
  T.~K\"ohler, K.~G\'oral, and P.~S.~Julienne,
  Rev.\ Mod.\ Phys.\ {\bf 78}, 1311 (2006).

\bibitem{Chin:2010}
  C.~Chin, R.~Grimm, P.~Julienne, and E.~Tiesinga,
  Rev.\ Mod.\ Phys.\ {\bf 82}, 1225 (2010).

\bibitem{de-Haas:1934}
  W.~J.~de~Haas, J.~de~Boer, and G.~J.~van~d\"en~Berg,
  Physica (Amsterdam) {\bf 1}, 1115 (1934).

\bibitem{Chevy:2010}
  F.~Chevy and C.~Mora,
  Rep.\ Prog.\ Phys.\ {\bf 73}, 112401 (2010).

\bibitem{Levinsen:2012}
  J.~Levinsen, P.~Massignan, F.~Chevy, and C.~Lobo,
  Phys.\ Rev.\ Lett.\ {\bf 109}, 075302 (2012).

\bibitem{Schirotzek:2009}
  A.~Schirotzek, C.-H.~Wu, A.~Sommer, and M.~W.~Zwierlein,
  Phys.\ Rev.\ Lett.\ {\bf 102}, 230402 (2009).

\bibitem{Kohstall:2012}
  C.~Kohstall, M.~Zaccanti, M.~Jag, A.~Trenkwalder, P.~Massignan, G.~M.~Bruun, F.~Schreck, and R.~Grimm,
  Nature (London) {\bf 485}, 615 (2012).

\bibitem{Knap:2012}
  M.~Knap, A.~Shashi, Y.~Nishida, A.~Imambekov, D.~A.~Abanin, and E.~Demler,
  Phys.\ Rev.\ X {\bf 2}, 041020 (2012).

\bibitem{Si:2010}
  Q.~Si and F.~Steglich,
  Science {\bf 329}, 1161 (2010).

\end{thebibliography}
\end{document}